\RequirePackage{fix-cm}
\documentclass[twocolumn]{svjour3}          
\smartqed  
\usepackage{dblfloatfix}
\usepackage{float}
\usepackage{graphicx}
\usepackage{lineno,hyperref}
\usepackage{times}
\usepackage{epsfig}
\usepackage{bm}
\usepackage{breakurl}
\usepackage{graphicx, subfigure}
\usepackage{amsmath,amssymb, amsfonts, euscript, mathrsfs}
\usepackage{subeqnarray}
\usepackage{cases}
\usepackage{enumitem}
\usepackage[sectionbib,round]{natbib}
\usepackage{algorithm}
\usepackage{algorithmic}
\usepackage{fancybox}
\usepackage{picins}
\usepackage{picinpar} 
\usepackage[table]{xcolor}
\usepackage{wrapfig}
\usepackage[numbered,framed]{matlab-prettifier}
\definecolor{lightgray}{gray}{0.9}
\definecolor{lightblue}{rgb}{0.98,0.98,1.0}
\usepackage[toc,page]{appendix}

\newcommand{\executeiffilenewer}[3]{%
\ifnum\pdfstrcmp{\pdffilemoddate{#1}}%
{\pdffilemoddate{#2}}>0%
{\immediate\write18{#3}}\fi%
}

\usepackage[normalem]{ulem}
\newcommand{\stkout}[1]{\ifmmode\text{\sout{\ensuremath{#1}}}\else\sout{#1}\fi}

\graphicspath{ {./Images/} }

\modulolinenumbers[5]

\journalname{Structural and Multidisciplinary Optimization}
\begin{document}\sloppy

\title{Stress Topology Analysis for Porous Infill Optimization
}


\author{Junpeng Wang         \and
        Jun Wu              \and
        R{\"u}diger Westermann              
}


\institute{J. Wang and R. Westermann \at
              Computer Graphics and Visualzation, Technical University of Munich, Munich, Germany \\
              \email{\{junpeng.wang, westermann\}@tum.de}           
           \and
           J. Wu \at
              Department of Sustainable Design Engineering, Delft University of Technology, Delft, The Netherlands \\
              Corresponding Author\\
              \email{j.wu-1@tudelft.nl}
}

\date{Received: date / Accepted: date}

\maketitle

\begin{abstract}
The optimization of porous infill structures via local volume constraints has become a popular approach in topology optimization. In some design settings, however, the iterative optimization process converges only slowly, or not at all even after several hundreds or thousands of iterations. This leads to regions in which a distinct binary design is difficult to achieve. Interpreting intermediate density values by applying a threshold results in large solid or void regions, leading to sub-optimal structures. We find that this convergence issue relates to the topology of the stress tensor field that is simulated when applying the same external forces on the solid design domain. In particular, low convergence is observed in regions around so-called trisector degenerate points. Based on this observation, we propose an automatic initialization process that prescribes the topological skeleton of the stress field into the material field as solid simulation elements. These elements guide the material deposition around the degenerate points, but can also be remodelled or removed during the optimization. We demonstrate significantly improved convergence rates in a number of use cases with complex stress topologies. The improved convergence is demonstrated for infill optimization under homogeneous as well as spatially varying local volume constraints.

\end{abstract}

\section{Introduction} \label{intro}
Topology optimization aims at finding the optimal structural layout under relevant design specifications. Topology optimization of multi-scale structures, which dates back to the seminal paper by~\cite{Bendsoe1988CMAME}, has been a topic of great interest in recent years. The rapid development in this field is partially stimulated by the possibility to fabricate complex structures using additive manufacturing. For an overview of topology optimization approaches for designing multi-scale structures, we refer readers to a recent review article by~\cite{Wu2021SMO}.

It has been shown that density-based topology optimization for compliance minimization, under local volume constraints, creates porous infill structures similar to those found in bone~\citep{Wu2018TVCG}. These bone-mimicking porous structures are lightweight, robust regarding material damages and loading variations, and stable with respect to buckling. The local volume constraints function similarly to maximum length scale control~\citep{Guest2009SMO}. They prevent the forming of large solid regions and, consequently, create porous structures distributed more evenly over the design domain. This approach has been extended, in conjunction with a coating approach proposed by~\cite{clausen2015topology}, to design concurrently structures and porous sub-structures therein, referred to as shell-infill composites~\citep{wu2017minimum}. It has also been applied to design porous shell structures~\citep{traff2021topology}. Other notable extensions include the design of porous structures with gradation in the porosity and pore size~\citep{schmidt2019structural,Das2020MD}, use of multiple materials~\citep{Li2020CMAME,Zhao2021SMO}, stress constraints~\citep{Kranz2021SMO}, as well as design of self-supporting infill~\citep{Liu2021SMO} and fiber-reinforced structures~\citep{Li2021CMAME}. \citeauthor{Dou2020SMO} proposed a projection operator to implicitly incorporate local volume constraints~\citep{Dou2020SMO}. Besides by density-based approaches, porous infill structures have been designed using an evolutionary design approach~\citep{Qiu2020AM} and machine learning~\citep{cang2019one}.

In this paper, we investigate the convergence behavior of density-based topology optimization with local volume constraints under a single load case. In density-based topology optimization, an important convergence criterion is that the optimized density field converges to a binary or so-called black-white design, i.e., the pseudo density is close to 1 or 0. A few hundred iterations or even more are not uncommon to achieve black-white designs. To improve the convergence rate, a typical solution is to apply a continuation scheme where parameters are updated after a certain number of iterations. However, in some optimization scenarios under local volume constraints, we have observed that certain regions fail to converge to a binary design even after thousands of iterations (see Fig.~\ref{fig:motivation}). Interpreting these intermediate density values by applying a threshold results in large solid or void regions, leading to sub-optimal structures.

To analyze the regions where low convergence is observed, we investigate the stress distribution in these regions via trajectory-based visualization~\citep{wang2020globally}. In particular, we shed light on the relationship between the convergence behavior and the principal stress directions that occurs when simulating on the solid design domain. This approach is inspired by previous work on infill optimization, where uniformly seeded tensor glyphs have confirmed good agreement between the optimized porous infill and the principal stress directions in the solid under load~\citep{Wu2018TVCG}. 
In this paper, we exploit advanced mechanisms to perform a topology-based analysis of the stress field, including the use of \textit{degenerate points} and \textit{topological skeletons}. At a degenerate point, the principal stress directions cannot be decided, yet a set of hyperbolic and parabolic sectors exist in its surrounding, in which similar patterns of neighboring trajectories are observed~\citep{delmarcelle1994topology}. The topological skeleton consists of the boundaries between adjacent sectors---so-called separatrices---and indicate pathways along which the forces are steered towards the degenerate points. In topology optimization, degenerate points have been used to indicate locations where integrability conditions are violated and consistent domain parameterizations cannot be computed~\citep{Stutz2020SMO}. 

When applying topology analysis to the stress tensor field, it reveals that low convergence occurs around a special type of degenerate points, known as trisectors. Notably, such degenerate points do not always appear, but if so, low convergence is often observed in their surrounding. Due to the isotropy of the stress tensor close to a trisector, the principal stress directions and, thus, a locally consistent binary material layout cannot be decided by the optimizer.
In our work, we propose an automatic pre-process that supports the optimizer in finding such a layout, resulting in significantly improved convergence rates in settings where trisectors are paramount. In particular, we build upon the efficient computation of degenerate point locations and separatrices, and prescribe an initial density field where elements along the separatrices are solid and all other elements take an intermediate value (i.e., the local volume upper bound). The solid elements along the separatrices can be changed over the course of the optimization to improve the structural performance, yet we observe that the optimization keeps them more or less unchanged and changes element densities in other parts accordingly. 
In the vicinity of trisectors, the initialization guides the optimizer towards a stable binary design and enables the optimization process to quickly converge towards a sound global layout. Interestingly, even though one might expect that the imposed initialization biases the optimizer towards a less stiff local optimum, the resulting binary designs show the same or even improved compliance compared to the designs generated by the original approach which exhibit unresolved intermediate density values in the presence of trisectors.  

The remainder of this paper is organized as follows. In Section~\ref{sec:PIO}, we first review the problem formulation underlying porous infill optimization. Then, in Section~\ref{sec:CAI} we analyse the convergence of porous infill optimization, elaborate on the relationships between optimization convergence and the existence of degenerate points in the stress field, and propose topology-based material initialization to counteract low optimization convergence. The implementation details of performing the topology analysis to the stress tensor field are discussed in Section~\ref{sec:impl}. We demonstrate the effectiveness of our approach in a variety of experiments in Section~\ref{sec:RD}. Section~\ref{sec:conclude} concludes the paper with a discussion of the proposed approach as well as future research directions. 


\section{Porous Infill Optimization} \label{sec:PIO}
The low convergence in some design tasks is observed while using the infill optimization approach~\citep{Wu2018TVCG}, on which and some of its extensions the effectiveness of our method will be demonstrated. For the sake of completeness, we briefly review the formulation of the density-based infill optimization with local volume constraints.

\subsection{Local volume constraints}
In a discretized design domain, the local volume ($\bar{\rho}_e$) of a circular region centered at the centroid of an element, $x_e$, is computed by 
\begin{equation} \label{eqn:calcLVF}
    \bar{\rho}_e = \frac{\sum_{i\in N_e}\rho_i}{\sum_{i\in N_e}1}, \quad N_e = \{i| \parallel x_i-x_e \parallel_2 \leq R_e \}, \:\: \forall e,
\end{equation}
where $\rho_i \in[0,1]$ is the pseudo density for the $i$-th element. $R_e$ denotes the radius of the region on which the local volume is measured.

An upper bound ($\alpha_e$, $0<\alpha_e<1$) is imposed on the local volume of each element in the design domain, i.e.,
\begin{equation} \label{eqn:LVF}
     \bar{\rho}_e \leq \alpha_e.
\end{equation}
Thus, the local volume constraint involves two parameters, $R_e$ and $\alpha_e$. $R_e$ indirectly controls the spacing between sub-structures, and $\alpha_e$ effectively controls the porosity~\citep{Wu2018TVCG}. In the original approach, both input fields are prescribed to be homogeneous. Recent developments have demonstrated the use of heterogeneous fields to generate gradations of the porosity and pore size of the optimized porous structures.

Assigning a local volume constraint to each element results in a large number of constraints that need to be considered by the optimizer. Dividing both sides of Eq.~\ref{eqn:LVF} by $\alpha_e$, these constraints are aggregated by the $p-$mean function,
\begin{equation} \label{eqn:finalLVF}
    \left(\frac{1}{n} \sum\limits_e \left(\frac{\bar{\rho}_e}{\alpha_e}\right)^p \right)^{\frac{1}{p}} \leq 1,
\end{equation}
where $n$ is the number of elements. $p=16$ is found to give a good approximation, and is used in this paper. 

\subsection{Optimization problem}
With the local volume constraint defined, the optimization problem is given by
\begin{align}
    \displaystyle \min \limits_{\bm{\phi}} \quad &  c=\frac{1}{2} \bm{U}^T \bm{K} \bm{U}, \label{eqn:obj}\\
    \displaystyle \mathrm{s.t.}  \quad & \bm{K} \bm{U} = \bm{F}, \label{eqn:FEA}\\
    \displaystyle & g \left(\bm{\phi} \right) = \left(\frac{1}{n} \sum\limits_e \left(\frac{\bar{\rho}_e}{\alpha_e}\right)^p \right)^{\frac{1}{p}} - 1 \leq 0, \label{eqn:cons}\\
    \displaystyle & \phi_e \in [0.0, \:\: 1.0], \:\: \forall e. \label{eqn:DD}
\end{align}

Here the objective is to minimize the compliance, measured by the strain energy $c$. $\bm{K}$ is the stiffness matrix in finite element analysis. $\bm{U}$ is the displacement vector, obtained by solving the static elasticity equation (Eq.~\ref{eqn:FEA}), where $\bm{F}$ is the loading vector. $g \left(\bm{\phi} \right)$ represents the aggregated local volume constraint. 

The formulation takes $\phi_e$ as the design variable. The pseudo density field ($\bm{\rho}$) is computed from $\bm{\phi}$ by a density filter ($\bm{\phi} \to \tilde{\bm{\phi}}$), followed by a smoothed Heaviside projection ($\tilde{\bm{\phi}} \to \bm{\rho}$). The density filter, with a filter radius $r_e$ smaller than $R_e$ (Eq.~\ref{eqn:calcLVF}), avoids checkerboard patterns resulting from numerical instabilities. The associated equations of this standard operator are omitted here but can be found in e.g.,~\citep{Wang2011SMO,Wu2018TVCG}. The purpose of the projection $\tilde{{\bm{\phi}}} \to \bm{\rho}$ is to promote a 0-1 solution, by thresholding at the value of $\frac{1}{2}$,
\begin{equation}
    \rho_e (\tilde{\phi}_e) = \frac{\tanh(\frac{\beta}{2})+\tanh(\beta(\tilde{\phi}_e-\frac{1}{2}))}{2\,\tanh(\frac{\beta}{2})} .
    \label{eq:softBlackWhiteFilter}
\end{equation}
The smoothed Heaviside function has a parameter, $\beta$, to control its sharpness. For improving convergence behaviour, a continuation scheme is applied to gradually increase its sharpness, i.e., we start with $\beta=1$ and double its value every 40 iterations until it reaches 128. 

To interpolate the Young's modulus for intermediate densities, we use the modified SIMP (Solid Isotropic Material with Penalization) model,
\begin{equation} \label{eqn:SIMP}
    E_e(\rho_e) = E_{min} + \rho_e^\gamma (E_0 - E_{min}),
\end{equation}
where $E_0$ is the Young's Modulus of a fully solid element. $E_{min}$ is a minimum Young's modulus ($E_{min}=1.0e^{-6}E_0$ in our test), introduced to avoid the singularity of the global stiffness matrix. $\gamma$ is the penalization factor, which is typically set to 3. $E_e(\rho_e)$ is the interpolated Young's Modulus of the element with density $\rho_e$.

The optimization problem is solved using the method of moving asymptotes (MMA)~\citep{svanberg1987method}. In all experiments performed in this work, the move limit of design variables is set to 0.01 unless specified otherwise. 
\begin{figure*}[t]
    \centering
    \includegraphics[width=0.98\linewidth, trim=0.0cm 3.0cm 1.5cm 0.0cm, clip=true]{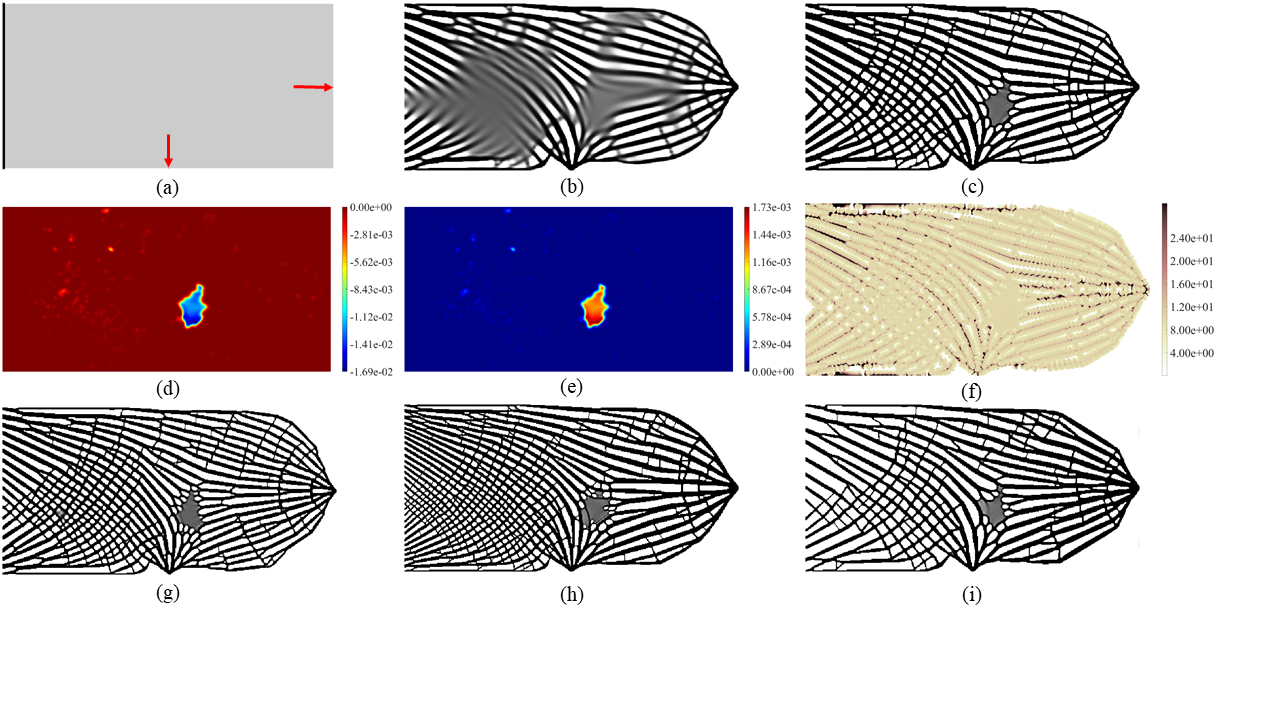}
    \caption{
    (a) Illustration of the design domain (500x250 simulation elements) and boundary conditions. (b, c) The density distributions after 250 and 1000 iterations, respectively, from topology optimization under local volume constraints. Design parameters are $\alpha_e=0.6$, $R_e=18$ and $r_e=4.5$. (d, e, f) The sensitivities at 1000 iterations, of the objective $\frac{\partial c}{\partial {\rho}}$, the constraint $\frac{\partial g}{\partial {\rho}}$, and $-\frac{\partial c}{\partial {\rho}} / \frac{\partial g}{\partial {\rho}}$.
    (g, h, i) show the optimized density fields under different parameter settings: (g) $R_e=12$. (h) $r_e=2.6$, $R_e$ varies linearly from $8$ to $24$, from the left to right side of the design domain.
    (i) $\alpha_e$ varies linearly from $0.4$ to $0.7$, from the left to right side of the design domain. All other parameters are kept the same as in (b) and (c).}
    \label{fig:motivation}     
\end{figure*}

\section{Convergence Analysis and Improvement} \label{sec:CAI}
When applying topology optimization using local volume constraints, in some scenarios it is observed that the iterative optimization process converges very slowly. When inspecting such scenarios in more detail, for instance, by visualizing the material distribution of the intermediate designs, it turns out that in some regions even after several hundreds or thousands of iterations a distinct binary design cannot be achieved by the optimizer. One of such scenarios is shown in Fig.~\ref{fig:motivation}. The rectangular design domain is fixed on its left edge. A unit load is applied on the right, while another unit load on the bottom, both in the middle of the edges. In Fig.~\ref{fig:motivation}(b), i.e., optimization after 250 iterations, two large grey regions can be observed. While the grey region on the left converges to a binary design after another 250 iterations, the grey region on the right does not result in a binary design even after a few thousands iterations. Applying a threshold to the intermediate densities to set them to either 0 or 1 results in large void or solid regions with sub-optimal mechanical properties or use of material.

To further analyze the cause of slow convergence, we examine the sensitivities at the 1000 iterations. In Fig.~\ref{fig:motivation}(d), the plot of $\frac{\partial c}{\partial {\rho}}$, the region where low convergence is observed has a high absolute sensitivity, meaning that an increase in density shall be favored for reducing the objective. However, an increase of density in this region will greatly violate the aggregated local volume constraint, as can be seen in Fig.~\ref{fig:motivation}(e), the plot of $\frac{\partial g}{\partial {\rho}}$. Shown in Fig.~\ref{fig:motivation}(f) is $-\frac{\partial c}{\partial {\rho}} / \frac{\partial g}{\partial {\rho}}$, a metric similarly used for deriving a fix-point type update scheme with the optimality criteria~\citep{Sigmund2001SMO}. It can be seen that in the low convergence region the ratio is rather homogeneous and does not indicate a clear material update strategy. 

\subsection{Relationship between Convergence and Stress}\label{sec:RCS}
Prior work in infill optimization has shown that the optimized porous structure is in many regions according to the principal stress directions that occur in the solid design domain under equal boundary conditions and external loads. 
At each point in a 2D solid under load, the stress state is fully described by the stress vectors for two mutually orthogonal orientations. The second-order stress tensor 
\begin{equation} \label{eqn:ST}
    S(x, y) = \begin{bmatrix} \sigma_{xx} & \tau_{xy} \\ \tau_{xy} & \sigma_{yy} \end{bmatrix} _{(x, y)}
\end{equation}
contains these vectors for the axes of a Cartesian coordinate system. 
$\sigma_{xx}$ and $\sigma_{yy}$ are the normal stress components along the $x$ and $y$ directions, respectively, $\tau_{xy}$ is the shear stress component. 

$S$ is symmetric since the shear stresses given by the off-diagonal elements in $S$ are equal on mutually orthogonal lines. 
The principal stress directions of the stress tensor indicate the two mutually orthogonal directions along which the shear stresses vanish. These directions are given by the eigenvectors of $S$, with magnitudes given by the corresponding eigenvalues $\sigma_1$ and $\sigma_2$ of $S$. 
For $\sigma_1 \geq \sigma_2$, $\sigma_1$ is called the major principal stress, and $\sigma_2$ the minor principal stress. Accordingly, the corresponding eigenvectors $v_1$ and $v_2$ are called major and minor principal stress directions. 
The signs of the principal stress magnitudes classify the stresses into tension (positive sign) or compression (negative sign). However, since there are two principal stresses acting at each point, the classification is with respect to a specific direction.

\begin{figure*}[t]
    \centering
    \includegraphics[width=0.98\linewidth, trim=0.0cm 12.5cm 2.0cm 0.0cm, clip=true]{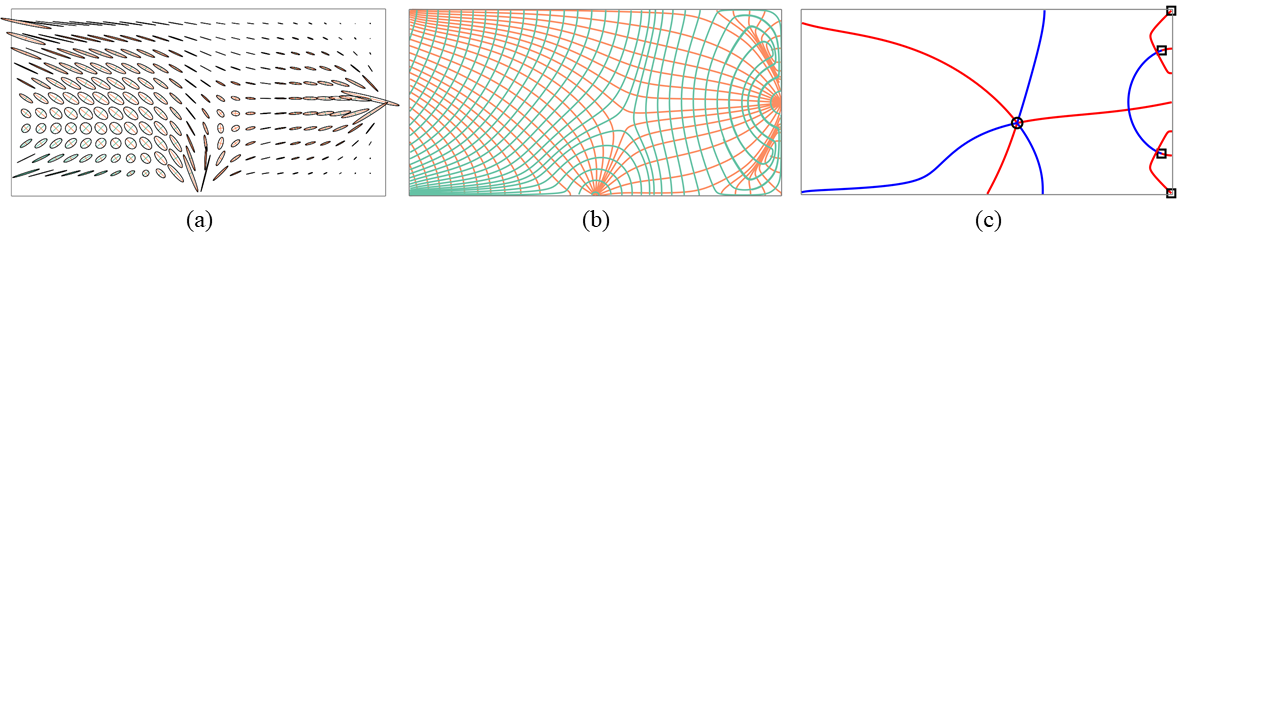}
    \caption{Stress tensor field visualizations. The stress tensor field is according to the scenario shown in Fig.~\ref{fig:motivation}a, when simulating on the solid design domain. (a) Tensor glyphs are drawn at sampled vertices of the Cartesian simulation grid. Colors indicate the sign of the principal stresses, red for positive and green for negative values. (b) Trajectory-based visualization. Orange and turquoise trajectories represent the major and minor principal stress directions, respectively. (c) Topology-based visualization. The circle and quads indicate the trisector and wedge degenerate points, respectively, which are connected via the topological skeleton. 
    }
    \label{fig:stressVis}      
\end{figure*}

Fig.~\ref{fig:stressVis}(a) shows a tensor glyph-based visualization of the stress field, corresponding to the scenario in Fig.~\ref{fig:motivation}. Here, the stress tensors are represented by oriented ellipses. The axes of the ellipses are oriented according to the eigenvectors of the stress tensor, and the lengths of their radii are determined by the eigenvalues. The colors of axes indicate the sign of the principal stresses, red for positive and green for negative values. Fig.~\ref{fig:motivation}(b), i.e., optimization after 250 iterations, has two large grey regions. Comparing it with Fig.~\ref{fig:stressVis}(a), it can be seen that the grey region on the left corresponds to $\sigma_1 = -\sigma_2$, and the one on the right corresponds to $\sigma_1 = \sigma_2$. While after a few hundred more iterations (see Fig.~\ref{fig:motivation}(c)) the grey region on the left converges to a binary design, the grey region on the right shrinks but remains visible.
In these regions, the optimizer can favour material growths either along the major or the minor principal stress direction, and it seems that because no preferential direction is present the optimizer has problems to decide for any of them. In the regions where the optimizer doesn't converge, however, another specific property can be perceived in addition to isotropic stress. As indicated by principal stress lines (PSLs), which are computed by performing numerical integration along the major and minor principal stress directions (see Fig.~\ref{fig:stressVis}(b)), these regions seem to cover locations where the PSLs indicate directional discontinuities in the tensor field. This observation gives rise to a stress topology-based analysis of the optimization convergence, which we provide in the following.

\subsection{Stress Topology-based Analysis}\label{sec:STA}

Topology analysis of 2D symmetric second-order tensor fields (e.g., stress tensor fields) has been introduced in the seminal work of  ~\cite{delmarcelle1994topology}. The topology of a 2D stress tensor field is composed of its \emph{degenerate points} and the corresponding \emph{topological skeleton}. At a degenerate point, the stress tensor has repeating eigenvalues, i.e., $\sigma_1 = \sigma_2$, meaning that the major and minor stress directions cannot be decided. The topological skeleton is given by principal stress lines---so-called separatrices---that start from degenerate points.

An isolated degenerate point can be classified by the winding number of one of the eigenvector fields on a loop surrounding the degenerate point, i.e., the tensor index. The most fundamental types of degenerate points are wedges and trisectors, with a tensor index of $1/2$ and $-1/2$, respectively. In Section~\ref{sec:impl}, we describe how the degenerate points are computed and classified for a stress tensor field given at the vertices of a Cartesian grid. A major/minor separatrix is a principal stress line starting at a degenerate point and following the major/minor eigenvector field. Let us also refer to Sec.~\ref{sec:impl} for a discussion of how to determine these directions. Fig.~\ref{fig:stressVis}c illustrates the major and minor separatrices in the stress field corresponding to the used test scenario in Fig.~\ref{fig:motivation}.

\begin{figure}
     \centering
     \includegraphics[width=0.98\linewidth, trim=0.0cm 1.5cm 7.0cm 0.0cm, clip=true]{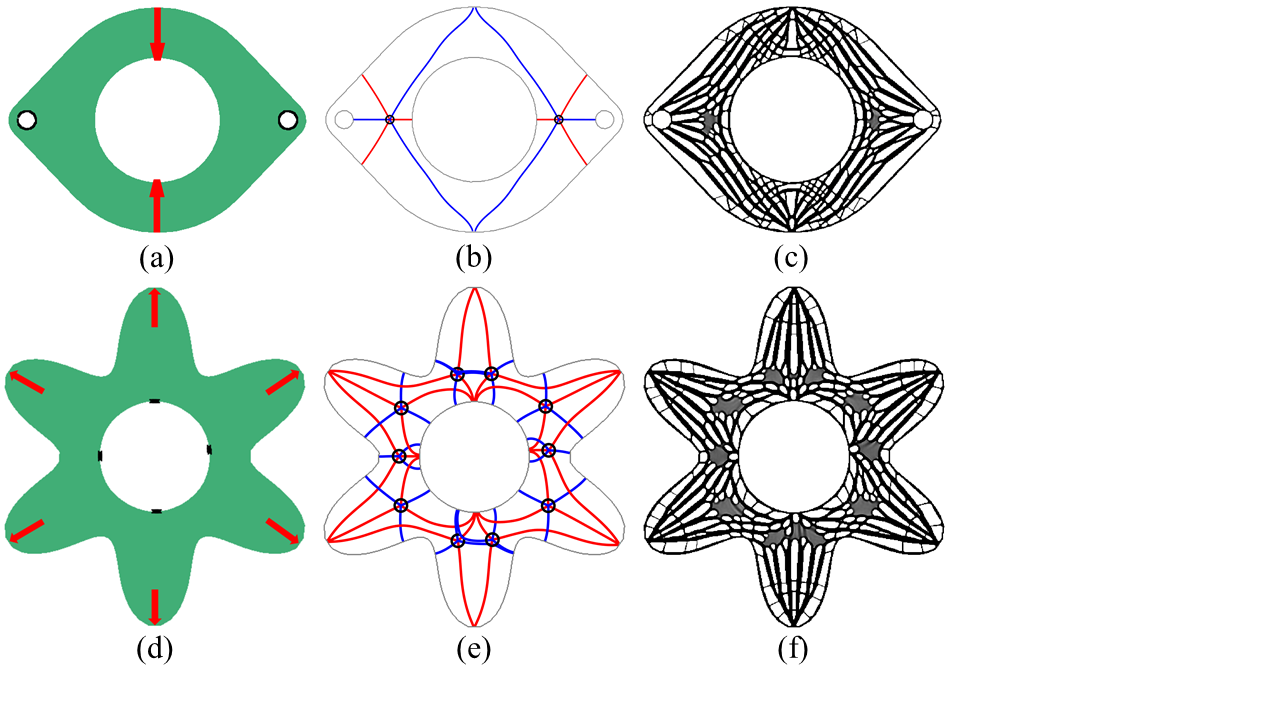} 
     \caption{Design domains and boundary conditions of models 'Bracket' (a) and 'Bearing' (d) using Cartesian simulation grids of resolutions $512\times400$ and $512\times512$, respectively. (b) (e): Trisector degenerate points and the corresponding topological skeletons. (c) (f): Density distributions after 1000 optimization iterations with $\alpha_e=0.6$, $R_e=18$ and $r_e=4.5$.}
     \label{fig:generalityConform}       
 \end{figure}

At a trisector there are three separatrices in the major (and three in the minor) principal direction field that divide the neighborhood into three sectors sharing this point. Around a wedge there can be either one sector or three sectors. In a sector, similar stress trajectories in the major and minor principal stress direction fields are observed. Fig.~\ref{fig:stressVis}c, as well as all other experiments we have performed, indicate that the regions where convergence cannot be achieved are always centered around a trisector, while wedges seem to have no influence on the convergence rate. Furthermore, regions where the convergence rate is low but the optimizer can eventually arrive at a stable binary design do not contain any degenerate point (Fig.~\ref{fig:motivation}b, left). These regions are characterized by isotropic stress, yet they are not affected by topological changes of the stress field.     
In Fig.~\ref{fig:generalityConform}, we give two more examples which verify our observations, and clearly indicate the relationships between convergence rate, stress isotropy, and existence of trisectors.

The topological skeletons can be perceived as limits of the principal stress lines close to the boundary between different stress regions. We hypothesize that the porous structures in regions where convergence is not achieved, if a stable binary design should be enforced, follow these skeletons, just as porous structures in other regions follow the principal stress directions. To validate this hypothesis, our idea is to guide the material deposition along the topological skeleton via a skeleton-based initialization of the density field. The initialization sets the optimizer to a state in which a stable binary design in the regions around trisectors is prescribed. The design can be changed during the course of the optimization, yet our experiments demonstrate that the optimizer maintains this design and builds additional support structures around it. These results empirically proof the validity of our hypothesis, and they indicate that the deviation from the prescribed skeleton is not favorable for the objective function.

\subsection{Stress Topology-guided Initialization} \label{sec:STI}
Typically in density-based topology optimization, the density field is initialized with a constant value. For infill optimization, the constant is chosen as the local volume upper bound. In accordance to the observation that the material layout in porous infill optimization is guided by the principal stress directions, we  propose to augment the initialization by setting the densities of elements close to the topological skeleton to a high value. This strategy is fully automatic, since the computation of neither the degenerate points nor the topological skeleton does involve any user intervention.

\begin{figure}
    \centering
    \includegraphics[width=0.42\linewidth]{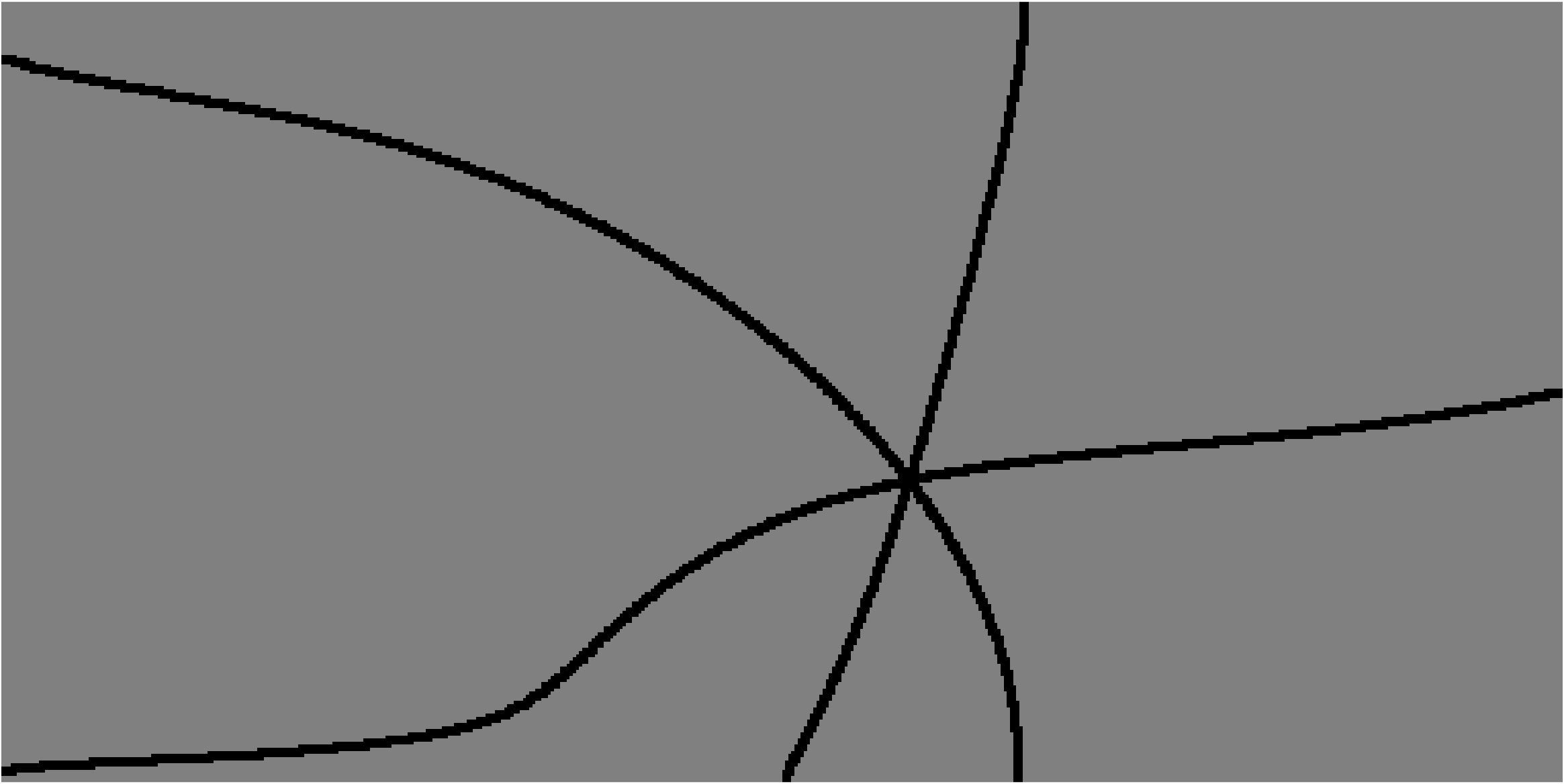}
    \includegraphics[width=0.26\linewidth]{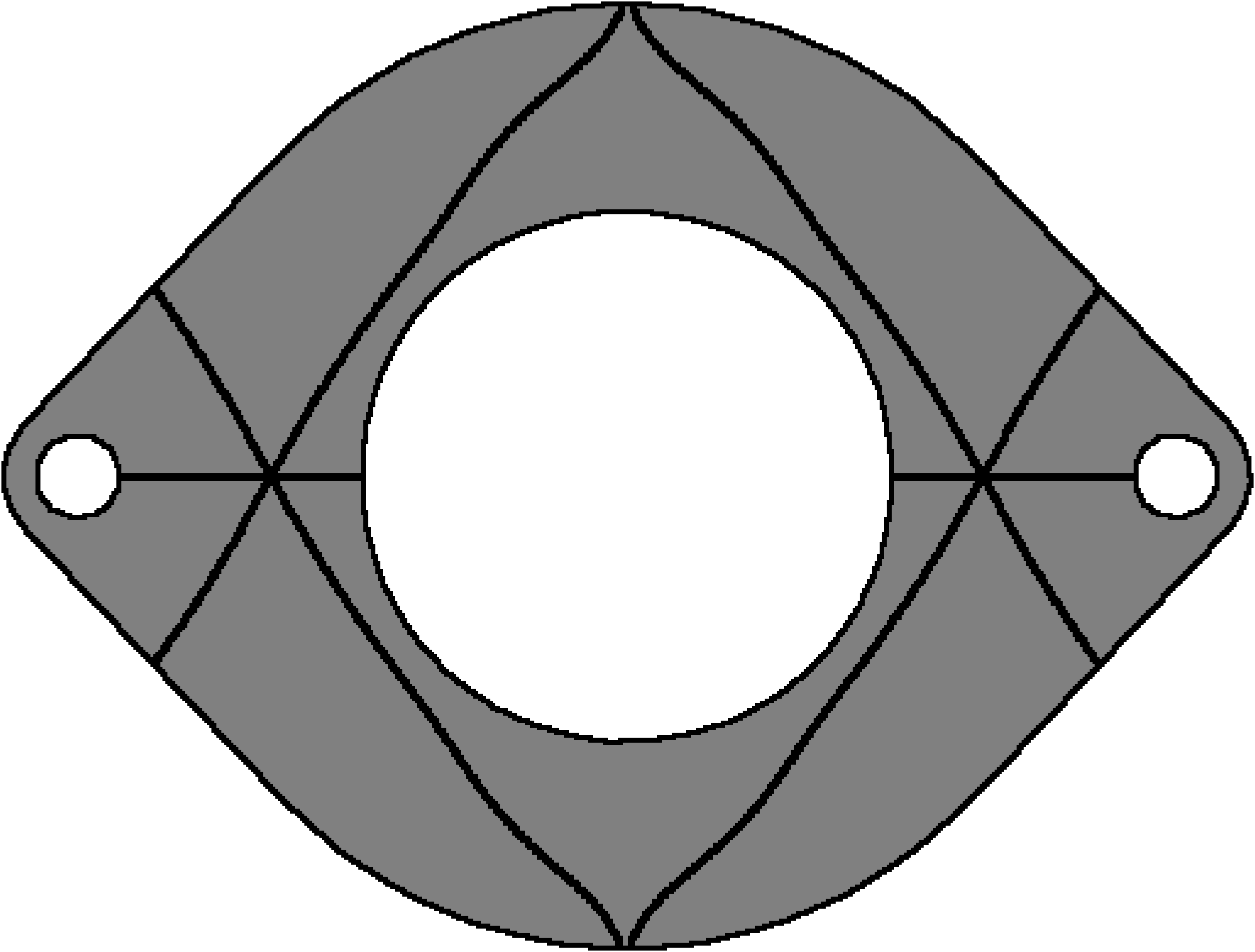}
    \includegraphics[width=0.22\linewidth]{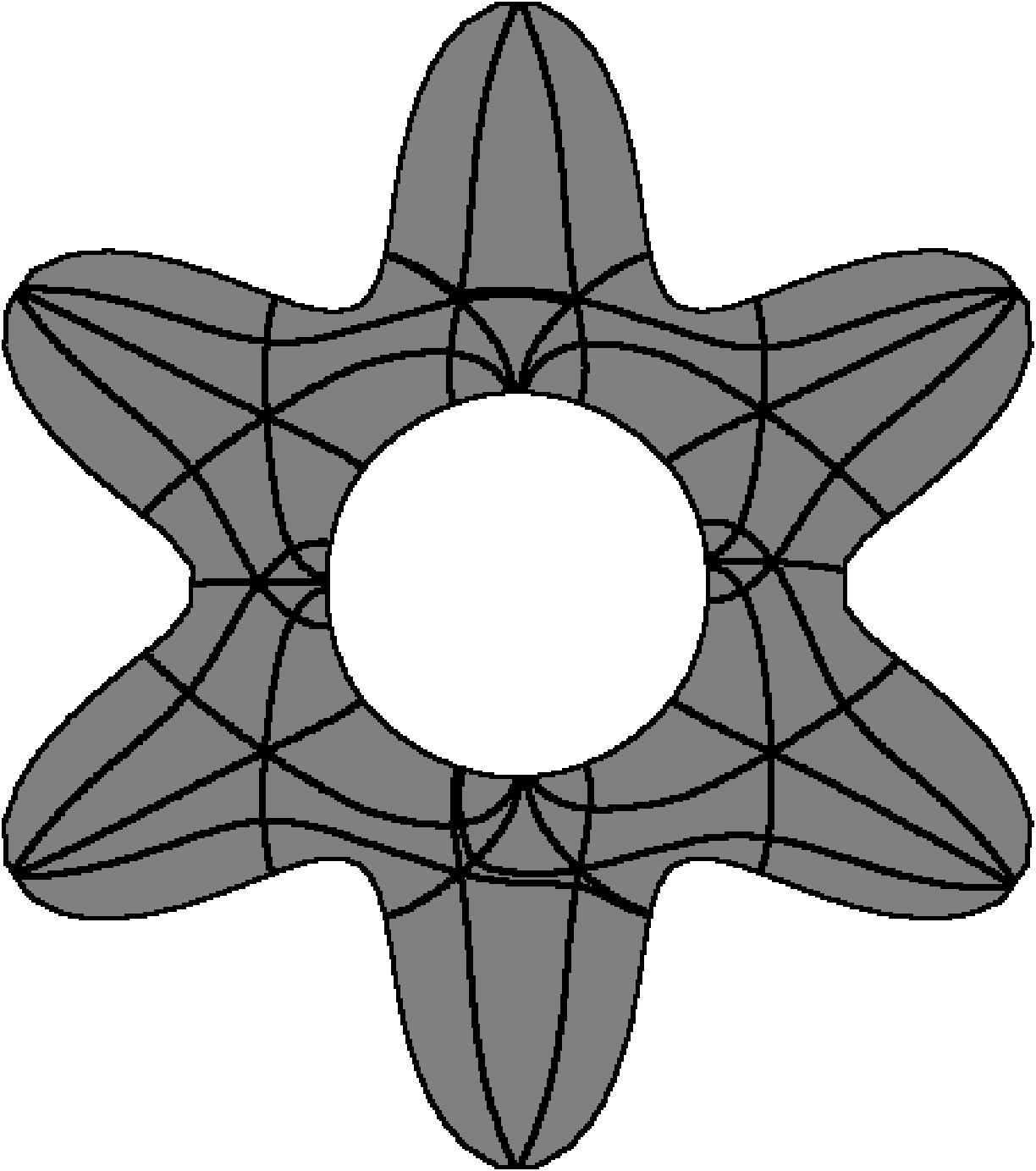}\\       
    \caption{The initialized material fields of the 'cantilever' in Fig.~\ref{fig:motivation} and 'Bracket' and 'Bearing' in Fig.~\ref{fig:generalityConform}.
    }
    \label{fig:Initialization}
\end{figure}

To generate the initial material layout, the elements which are near the topological skeleton are identified first. In the current implementation, all elements that are touched by any of the PSLs belonging to the skeleton are identified. Then, the initial volume fraction of these elements is set to solid at the beginning of the optimization process. In this way, the initial material field around a trisector degenerate point becomes inhomogeneous, giving rise to sensitivities favoring a unique topology layout. It is worth noting that these pre-embedded solid elements are not passive elements but still belong to the design space, i.e., the material density at these elements can be adjusted by the optimizer if a stiffer design can be achieved. Fig.~\ref{fig:Initialization} shows the initial material fields that are used in the test cases 'Cantilever', 'Bracket' and 'Bearing'.

The proposed initialization process can be integrated into porous infill optimization in a fully automated way. Once the design domain, material parameters, fixations and external load conditions are given, the following steps are performed:

\begin{itemize}
    \item [1)] Finite element analysis to compute the stress field in the fully solid design domain.
    \item [2)] Topology analysis including the computation of all trisector degenerate points and the topological skeleton containing these points.
    \item [3)] Initialization of the material field according to the topological skeleton. 
    \item [4)] Topology optimization using local volume constraints for porous infill optimization.
\end{itemize}

\section{Implementation Details} \label{sec:impl}
In the following, we discuss the computation of the locations of degenerate points in a given Cartesian simulation grids, as well as the computation of the topological skeleton that is required to initialize the material field. 
Given the definition of degenerate points, a degenerate point can be located by solving the following system of equations:
\begin{equation} \label{eqn:degeCond}
    \begin{array}{l}
        \sigma_{xx}(x^*, y^*)-\sigma_{yy}(x^*, y^*) = 0, \\
        \tau_{xy}(x^*, y^*) = 0,
    \end{array}
\end{equation}
where $(x^*, y^*)$ denotes the coordinates of the point to be solved for. Here we consider the general situation in topology optimization, i.e., the finite element analysis is performed using axis-aligned quadrilateral finite elements with bilinear shape functions. Thus, each element has four nodes that coincide with the element’s vertices, and the values at the nodes are bilinearly interpolated within the element. Then, Eq.~\ref{eqn:degeCond} becomes a non-linear system of equations, which can be solved by the Newton-Raphson method. 

\begin{figure}
    \centering
    \includegraphics[width=0.8\linewidth]{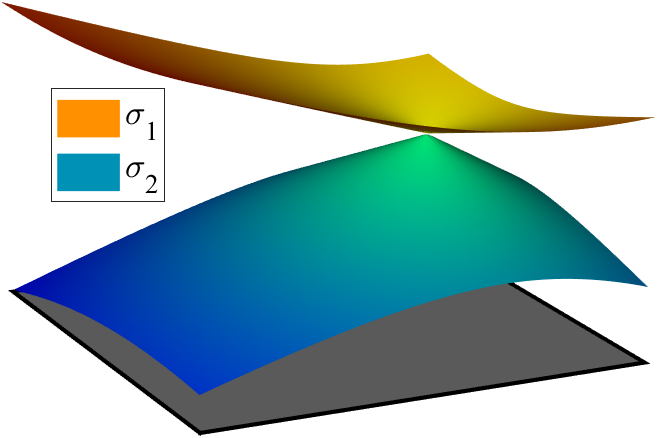}
    \caption{The eigenvalues corresponding to the major ($\sigma_1$) and minor ($\sigma_2$) principal stress direction are shown as height fields over the domain of a simulation element (grey square). At a degenerate point, both eigenvalues have the same value.}
    \label{fig:degePotSche}      
 \end{figure}

Since degenerate points usually appear only in a few elements, an efficient way is required to test whether a cell can contain such a point and needs to be further analysed, or can be excluded right away. Therefore, each element is first classified according to the following conditions: 

\begin{equation} \label{eqn:discriminant}
    \begin{array}{l}
        \sigma_{xx}(x_i,y_i)-\sigma_{yy}(x_i,y_i) > 0, \quad i=1:4 \quad \mathrm{or} \\
        \sigma_{xx}(x_i,y_i)-\sigma_{yy}(x_i,y_i) < 0, \quad i=1:4 \quad \mathrm{or} \\
        \tau_{xy}(x_i,y_i) > 0, \quad i=1:4 \quad \mathrm{or} \\        
        \tau_{xy}(x_i,y_i) < 0, \quad i=1:4
    \end{array}
\end{equation}
where $(x_i, y_i), \:\: i=1:4$ refers to the four nodal coordinates of a finite element. It can be easily shown that an element cannot contain a degenerate point if any of the conditions in Eq.~\ref{eqn:discriminant} is true. If none of the conditions is true, the element needs to be further analyzed to locate a degenerate point in its interior.  Fig.~\ref{fig:degePotSche} shows a possible distribution of the eigenvalues corresponding to the major and minor principal stress directions in a quadrilateral simulation element containing a degenerate point.

In a symmetric second tensor field, two types of stable degenerate points exist: trisectors and wedges. They are indicated by characteristic patterns of the PSLs in their vicinity, and are determined from the so-called tensor gradients (see \cite{delmarcelle1994topology} for a comprehensive derivation). First, the partial derivatives of the tensor are introduced as 
\begin{equation} \label{eqn:partialDerivative}
    \begin{array}{l}
        a = \frac{1}{2} \frac{\partial {(\sigma_{xx}-\sigma_{yy})}}{\partial x} \quad b = \frac{1}{2} \frac{\partial {(\sigma_{xx}-\sigma_{yy})}}{\partial y} \\
        c = \frac{\partial {\tau_{xy}}}{\partial x} \quad \quad \quad \quad \:\: d = \frac{\partial {\tau_{xy}}}{\partial y} 
    \end{array}
\end{equation}
These derivatives are then used to compute the invariant under rotation 
\begin{equation} \label{eqn:invarRotat}
    \delta = ad - bc.
\end{equation}

The sign of $\delta$ determines the type of the degenerate point. I.e., a trisector degenerate point is indicated by $\delta < 0$, and a wedge degenerate point is indicated by $\delta > 0$. At a trisector degenerate point, there are three major and three minor separatrices starting from this point. In contrast, two separatrices start from a wedge, one coincides with the major PSL and the other one with the minor PSL (see Fig.~\ref{fig:stressVis}c). These separatrices are termed the topological skeleton of a stress tensor field, i.e., the topological skeleton is composed of the PSLs starting from the degenerate points. Compared to the PSLs not belonging to the topological skeleton, the tangent of the topological skeleton at the degenerate point is not unique, since there is an infinite set of principal stress directions at such points. To solve this problem, \cite{delmarcelle1994topology}, propose that the tangents to the topological skeleton at the degenerate points are the real root(s) of the cubic equation 
\begin{equation} \label{eqn:cubicEq}
    dx^3 + (c+2b)x^2 + (2a-d)x - c = 0.
\end{equation}

\section{Results and Discussions} \label{sec:RD}
In this section, we use several examples to demonstrate the effectiveness of the proposed initialization for density-based porous infill optimization. All design domains are discretized by Cartesian finite element grids with unit size simulation elements. The Young's Modulus and Poisson's ratio are set to 1.0 and 0.3, respectively. Convergence improvement is quantified by the sharpness measurement 
\begin{equation} \label{eqn:sharp}
    s = \frac{4}{n} \sum\limits_{e} \rho_e(1-\rho_e)
\end{equation}
A small value of $s$ indicates a sharper binary design of the optimized topology. 

\begin{figure*}
    \centering
    \includegraphics[width=0.98\linewidth, trim=0.0cm 14cm 0.5cm 0.0cm, clip=true]{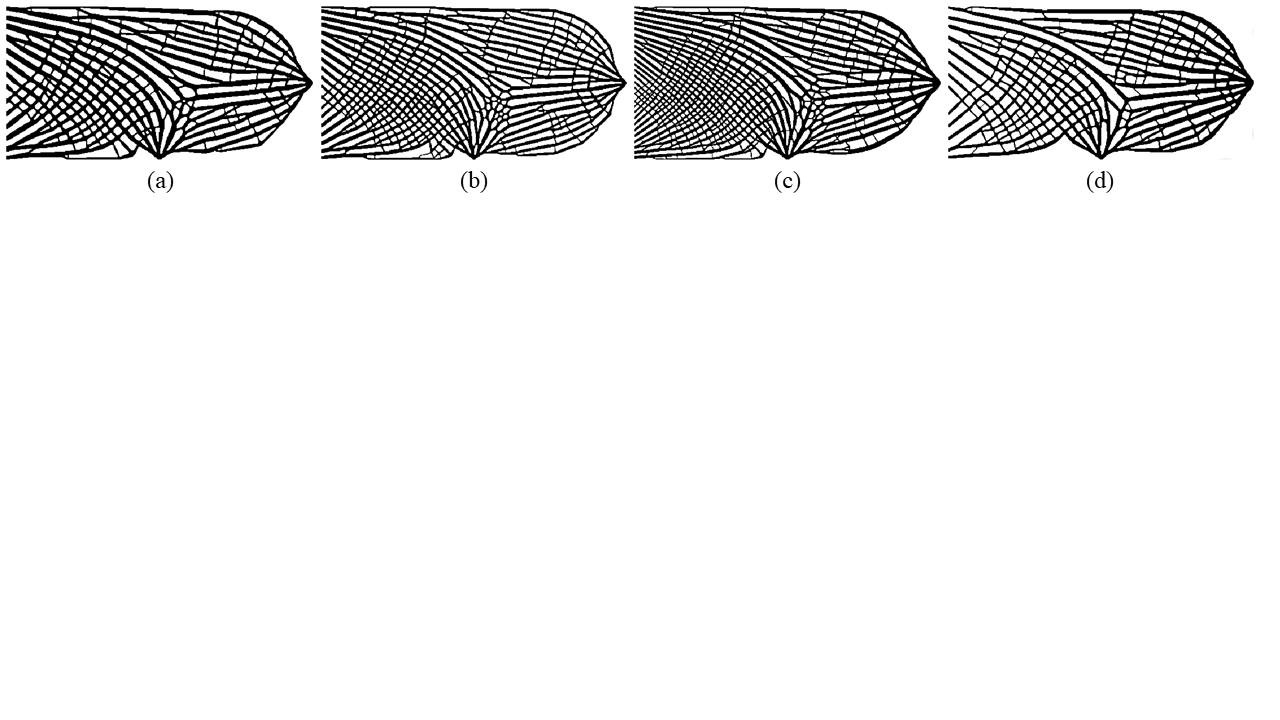}
    \caption{Binary designs by porous infill optimization with stress topology-guided material initialization. The same parameter settings as in Fig.~\ref{fig:motivation}c), g), h) and i), respectively, are used. A quantitative comparison is given in Table ~\ref{tab:qualitativeStatistics}. 
    }
    \label{fig:firstWorked}      
\end{figure*}

Figure~\ref{fig:firstWorked} shows the binary designs that are generated using an initialization by topological skeletons. The same parameter settings as in Fig.~\ref{fig:motivation} are used here. As can be seen, in all cases a binary design is achieved regardless of the area of the region around the degenerate point where convergence is not achieved by the original approach. 
Table~\ref{tab:qualitativeStatistics} compares the mechanical properties of the designs generated by both approaches, as well as the used material and the sharpness (cf. Eq.~\ref{eqn:sharp}) of the designs after 1000 optimization iterations. Notably, even after some thousands of iterations convergence cannot be reached via original porous infill optimization. As can be seen from the sharpness values, the proposed initialization strategy improves the convergence behavior of porous infill optimization considerably. In all test cases, a distinct binary design has been reached within the given number of iterations. The difference in compliance and material fraction is rather small.

\begin{table}
\centering
\caption{Quality and convergence comparison. Each pair of rows shows the compliance, the fraction of solid material, and the sharpness of the resulting designs when using porous infill optimization without (top) and with (bottom) stress topology-guided material initialization.}
\label{tab:qualitativeStatistics}       
\begin{tabular}{c|ccc}
\hline\noalign{\smallskip}
Cases & Compliance & Solid fraction & Sharpness\\
\noalign{\smallskip}\hline\noalign{\smallskip}
Fig.~\ref{fig:motivation}c & 27.99 & 0.479 & $1.8\times10^{-2}$ \\
Fig.~\ref{fig:firstWorked}a & 28.26 & 0.473 & $5.4\times10^{-3}$ \\
\hline
Fig.~\ref{fig:motivation}g & 27.37 & 0.501 & $1.1\times10^{-2}$ \\
Fig.~\ref{fig:firstWorked}b & 27.63 & 0.497 & $1.1\times10^{-3}$ \\
\hline
Fig.~\ref{fig:motivation}h & 35.49 & 0.377 & $1.0\times10^{-2}$ \\
Fig.~\ref{fig:firstWorked}c & 35.76 & 0.379 & $5.4\times10^{-3}$ \\
\hline
Fig.~\ref{fig:motivation}i & 27.98 & 0.482 & $1.2\times10^{-2}$ \\
Fig.~\ref{fig:firstWorked}d & 28.47 & 0.480 & $1.3\times10^{-3}$ \\
\hline
\hline
Fig.~\ref{fig:generalityConform}c & 19.15 & 0.490 & $1.7\times10^{-2}$ \\
Fig.~\ref{fig:ClassicPIresult}a & 19.32 & 0.482 & $5.8\times10^{-3}$ \\
\hline
Fig.~\ref{fig:generalityConform}f & 45.67 & 0.515 & $6.6\times10^{-2}$ \\
Fig.~\ref{fig:ClassicPIresult}b & 45.49 & 0.504 & $4.8\times10^{-3}$ \\
\noalign{\smallskip}\hline
\end{tabular}
\end{table}

\begin{figure}
    \centering
    \includegraphics[width=0.98\linewidth, trim=0.0cm 12cm 20.0cm 0.0cm, clip=true]{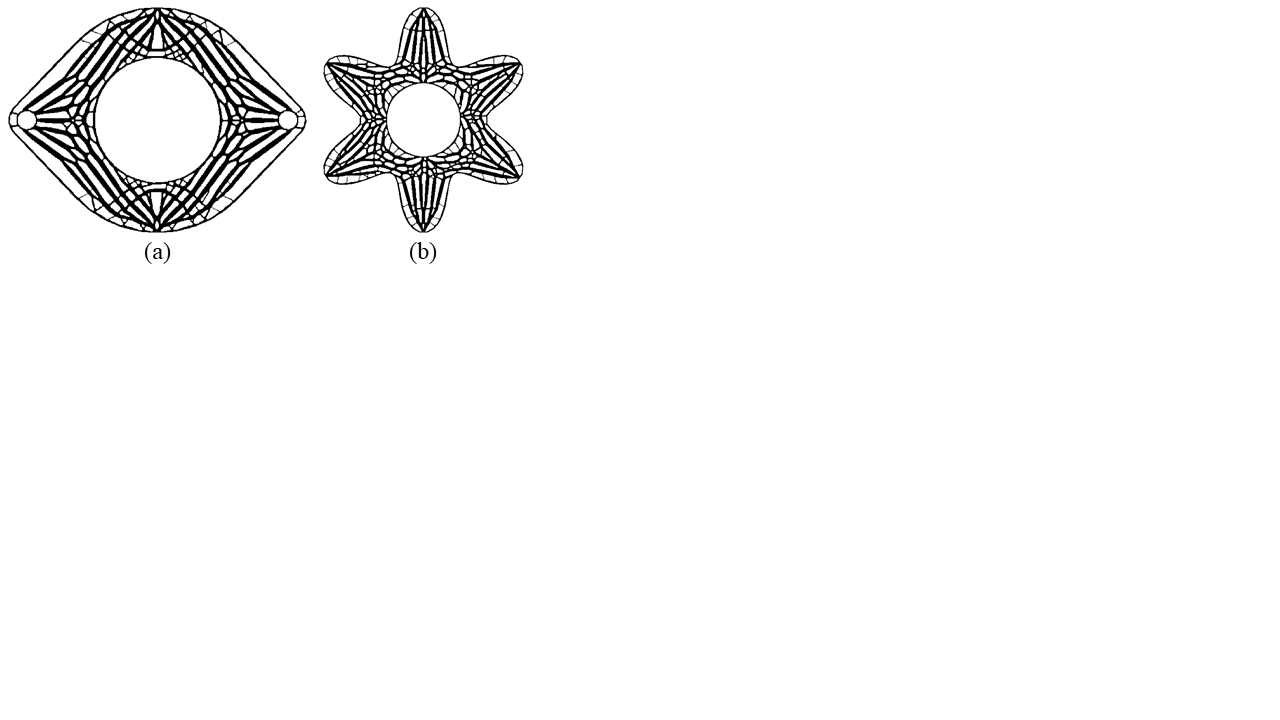}  
    \caption{
    The binary designs that are generated by stress topology-guided porous infill optimization for 'Bracket' (a) and 'Bearing' (b) from Fig.~\ref{fig:generalityConform}.}
    \label{fig:ClassicPIresult}       
\end{figure}

Figure~\ref{fig:ClassicPIresult} shows converged density distributions of the 'Bracket' and 'Bearing', obtained using the topology-based initialization strategy. Low convergence regions from the original approach (cf. Fig.~\ref{fig:generalityConform}) are removed. The convergence is again confirmed by a reduction in the sharpness value. In the 'Bearing' result from the original approach (cf. Fig.~\ref{fig:generalityConform}f) the area of the low convergence regions is large. In this case, the sharpness value is reduced by an order of magnitude by the proposed initialization. It leads to a stiffer structure with less material consumption.

\begin{figure}
    \centering
    \includegraphics[width=0.98\linewidth, trim=0.0cm 0.5cm 16cm 0.0cm, clip=true]{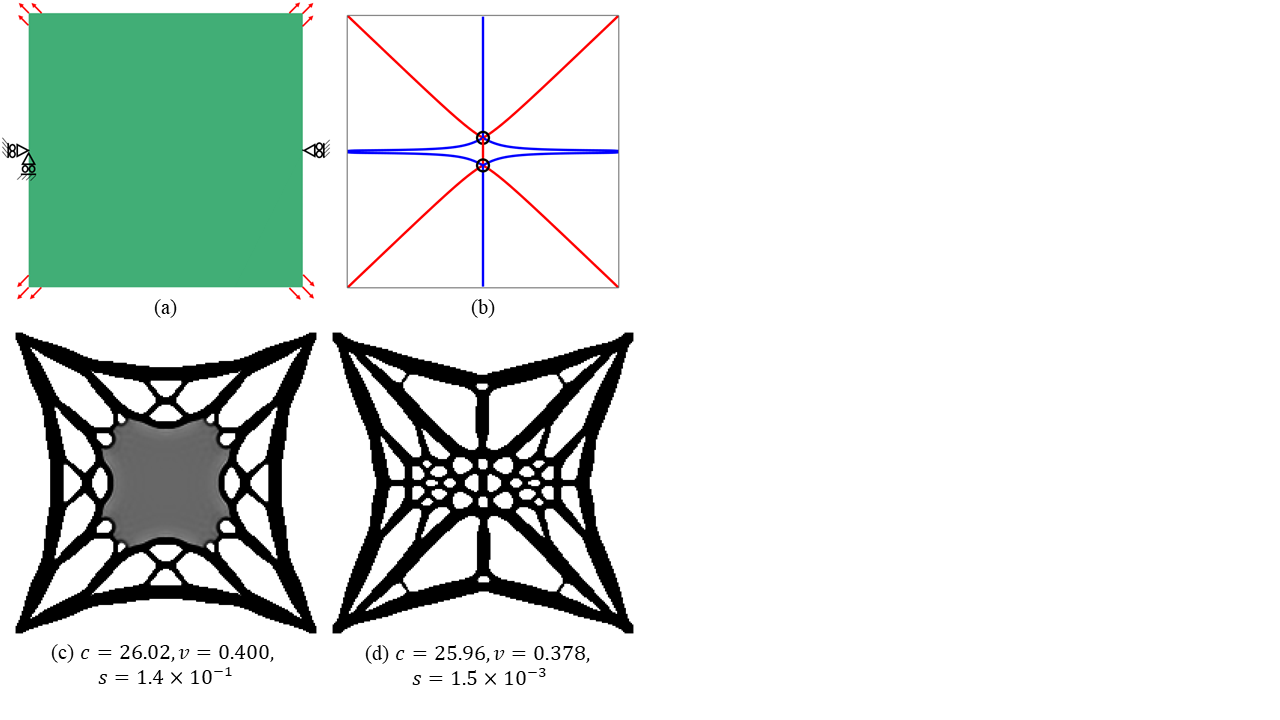}
    \caption{Porous infill optimization with both local and global volume fraction constraints. The optimization settings are $\alpha_e=0.6$, $\alpha_{total}=0.4$, $R_e=18$, $r_e=4.5$ and 1000 iterations. (a) The design domain ($200\times200$) and boundary conditions. (b) Trisector degenerate points and the corresponding topological skeletons. (c) The density distribution generated by porous infill optimization, and (d) using the proposed topology-guided material initialization.}    
    \label{fig:HybridPIresult}       
\end{figure}

\begin{figure}
    \centering
    \includegraphics[width=0.49\linewidth]{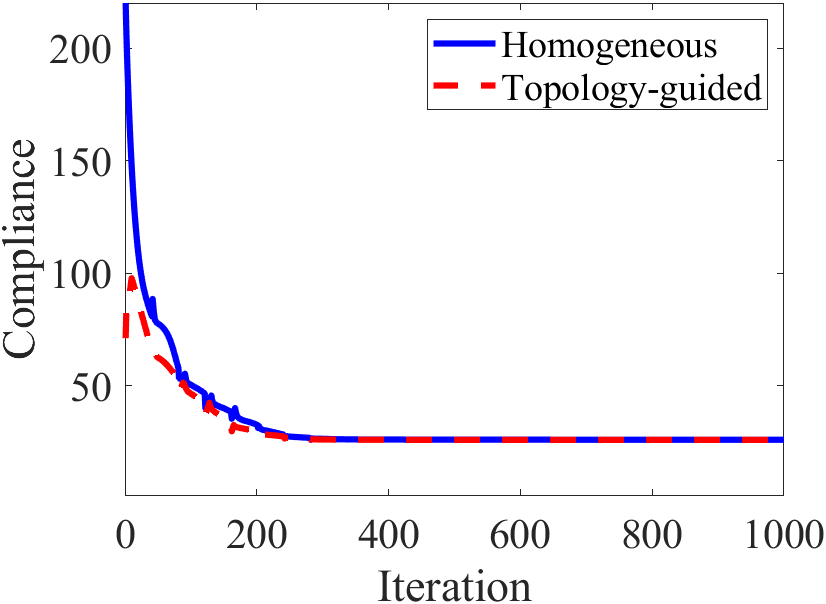}
    \includegraphics[width=0.49\linewidth]{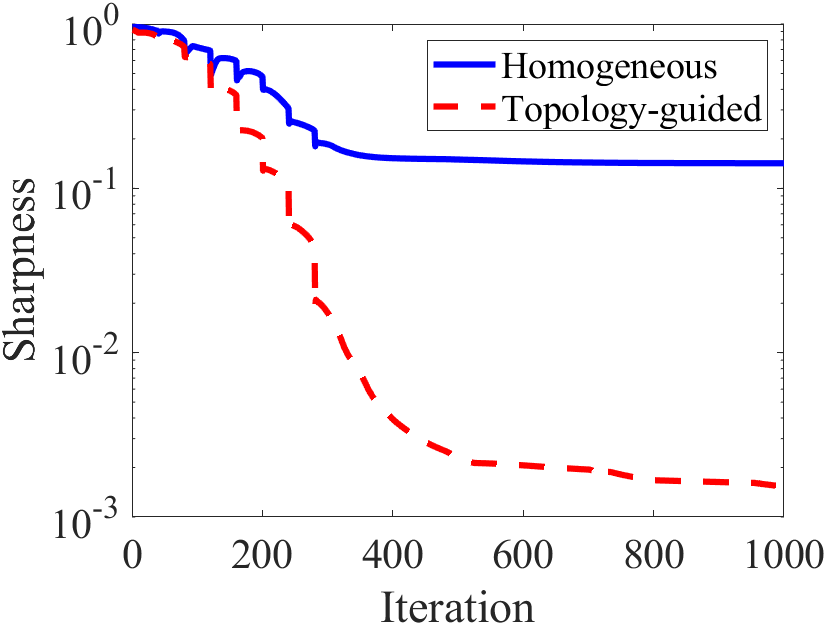}    
    \caption{Convergence plots for the example shown in Fig.~\ref{fig:HybridPIresult}, comparing the effects of homogeneous initialization and topology-guided initialization regarding the objective (left) and sharpness (right).}
    \label{fig:HybridPIstatistics}       
\end{figure}

We further test the applicability of the proposed initialization on topology optimization with both local and global volume constraints. The global volume constraint is 
\begin{equation} \label{eqn:gVFC}
    \frac{1}{n} \sum\limits_e \rho_e - \alpha_{total} \leq 0.
\end{equation}
The test is performed on a square where its four corners are loaded (Fig.~\ref{fig:HybridPIresult}(a)), an example taken from~\cite{Stutz2020SMO}. This example has two trisectors, as shown in Fig.~\ref{fig:HybridPIresult}(b). From Fig.~\ref{fig:HybridPIresult}(c), it can be seen that the central region is largely grey after 1000 optimization iterations. The grey region disappears in the optimized result from the proposed initialization (Fig.~\ref{fig:HybridPIresult}(d)). The significant improvement in convergence can also be seen from the plot of the sharpness over iterations, shown in Fig.~\ref{fig:HybridPIstatistics}(right). As the large grey region is replaced by a binary design, the material consumption reduces from 0.400 to 0.378 and the compliance value decreases marginally from 26.02 to 25.96.

\begin{figure}
    \centering
    \includegraphics[width=0.98\linewidth, trim=0.0cm 2.0cm 7.0cm 0.0cm, clip=true]{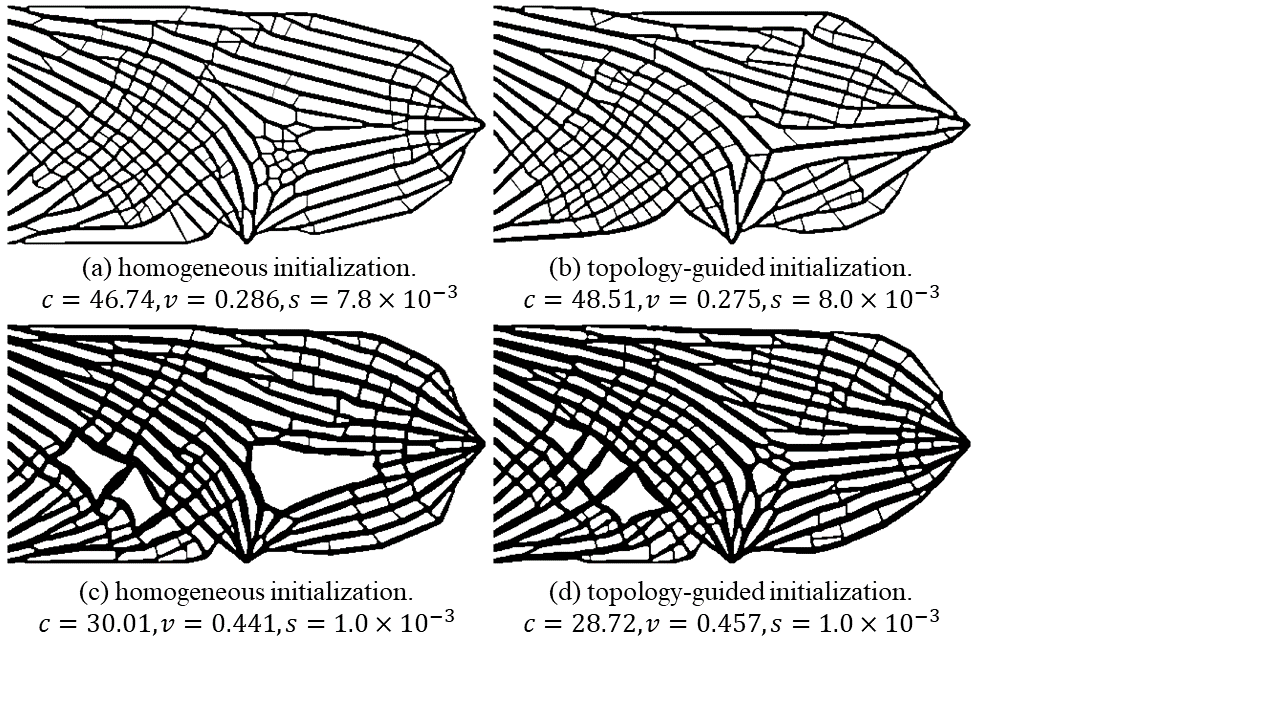}
    \caption{
    Density distributions of special cases of 'Cantilever' after 1000 iterations. Control parameters are kept the same as in Fig.~\ref{fig:motivation}c if not stated otherwise. (a) Using a homogeneous initial material field and $\alpha_e = 0.4$. (b) Using a stress topology-guided initial material field and $\alpha_e = 0.4$. (c) Using a homogeneous initial material field and setting the moving limit of MMA to 0.1. (d) Using a stress topology-guided initial material field and setting the moving limit of MMA to 0.1. 
    }
    \label{fig:sideSupport}       
\end{figure}

Interestingly, under certain design specifications, the original approach is able to create binary designs at the presence of degenerate points. This happens if the specified local volume bound is small. Fig.~\ref{fig:sideSupport} shows the cantilever example optimized under $\alpha_e=0.4$, with a homogeneous initialization (a) and with the proposed topology-guided initialization (b). The sharpness values of results from both initialization strategies are very close. The one with the proposed initialization leads to a slightly larger compliance ($103.8\%$), while consuming $3.8\%$ less material. Our experiments also revealed that the local convergence may, to some extend, be alleviated by a more aggressive move limit in the MMA solver. Fig.~\ref{fig:sideSupport}(c) shows the optimized result with a move limit of 0.1 (in contrast to a limit of 0.01 in previous examples), under a homogeneous initialization. While a binary design is obtained, the design has irregular large void regions that do not agree with the intention of creating distributed porous infill structures. The introduction of the topology-guided initialization is able to fill the void (see Fig.~\ref{fig:sideSupport}(d)). The latter design consumes $3.6\%$ more material, and decreases the compliance by $4.3\%$.

\section{Conclusions} \label{sec:conclude}
In this work, we have analyzed the convergence of porous infill optimization towards a stable binary design. In a number of experiments we have shown that low convergence regions may appear in this variant of topology optimization, prohibiting an automatic generation of a distinct and mechanically sound binary design. By analyzing the topology of the stress field that arises in the solid object, the existence of trisector degenerate points in this field could be determined as the major cause of low convergence. Based on this observation, we have proposed an initialization process for porous infill optimization that quickly guides the optimization towards a stable binary design. This process generates an initial solid material layout along the topological skeleton of the stress field, which is comprised of principal stress lines starting at the trisector degenerate points.   

In the future, we intend to shed light on the following extensions of the proposed approach: Firstly, we aim to consider the application of stress topology-based porous infill optimization to three-dimensional (3D) domains. Therefore, the convergence of 3D porous infill optimization first needs to be analyzed, using dedicated visualization techniques for 3D scalar fields. Then, since degenerate points become lines and surfaces in 3D, the relationships between the 3D stress field topology and the local convergence ratio needs to be investigated. Based on these investigations, specific initialization strategies and material growth processes need to be developed. Secondly, we will consider stress topology analysis for homogenization-based infill optimization. In particular, we will address the automatic generation of a 2D quad-dominant mesh where the mesh edges align with the principal stress directions. Porous infill optimization, under a single load case, tends to lay out the material along the mutually orthogonal principal stress lines, and---with our proposed initialization---automatically handles the material layout around degenerate points where quad meshing approach have difficulties to construct a consistent mesh structure~\citep{Wu2021TVCG}. We will build upon this observation and combine stress topology-guided porous infill optimization with the enforcement of material deposition along the principal stress lines.

\vspace{2mm}
\noindent
\small{\textbf{Acknowledgements} This work was supported in part by a grant from German Research Foundation (DFG) under grant number WE 2754/10-1.}

\vspace{2mm}
\noindent
\small{\textbf{Conflict of interest} The authors declare that they have no conflict of interest.}

\vspace{2mm}
\noindent
\small{\textbf{Replication of results} All important details have been presented in the paper. To facilitate re-implementation and reuse of this method, we provide a demonstration program\footnote{https://github.com/Junpeng-Wang-TUM/Infill$\_$plus} in Matlab. This demo works with rectangular design domains (i.e., Fig.~\ref{fig:motivation} and Fig.~\ref{fig:HybridPIresult}). This demo additionally includes the functionality for topology analysis and visualization of the stress tensor field. }

%
%

\bibliographystyle{spbasic}      
\bibliography{ref}   




\end{document}